\documentclass[10pt,conference]{IEEEtran}
\usepackage{color,soul}
\sethlcolor{yellow}
\usepackage{amsmath, amssymb, bm, cite, epsfig, psfrag}
\usepackage{graphicx}
\usepackage[top    = 0.7in,
bottom = 0.95in,
left   = 0.7in,
right  = 0.7in]{geometry}
\usepackage{array}
\usepackage{textcomp}
\usepackage{makecell}
\usepackage{multirow}
\newcommand{\tabitem}{~~\llap{\textbullet}~~}
\usepackage{longtable}
\usepackage{supertabular,booktabs}
\usepackage{multirow}
\usepackage[usenames,dvipsnames]{xcolor}

\usepackage{etoolbox}
\usepackage{pbox}
\usepackage{romannum}
\usepackage{enumerate}
\graphicspath{ {figures/} }
\usepackage{subfigure}
\usepackage{colortbl}
\usepackage{fancyhdr}
\usepackage{bm}

\graphicspath{{figures/}}

\usepackage{etoolbox}
\usepackage{fancyhdr}

\usepackage{bm}
\newtoggle{conference}
\togglefalse{conference} %
\graphicspath{{figures/}}

\fancypagestyle{firststyle}{
	\fancyhf{}
	\fancyhead[L]{O. Kanhere, H. Poddar, Y. Xing, D. Shakya, S. Ju, and T. S. Rappaport, ``A Power Efficiency Metric for Comparing Energy Consumption in Future Wireless Networks in the Millimeter-Wave and Terahertz bands,'' \textit{in IEEE Wireless Communications} doi: 10.1109/MWC.005.2200083 URL: https://ieeexplore.ieee.org/document/9864328 }    %

}

\begin{document}

\title{A Power Efficiency Metric for Comparing Energy Consumption in Future Wireless Networks in the Millimeter-Wave and Terahertz bands}%
\author{\IEEEauthorblockN{Ojas Kanhere, Hitesh Poddar, Yunchou Xing, Dipankar Shakya, Shihao Ju, and Theodore S. Rappaport\\}
	
\IEEEauthorblockA{	\small NYU WIRELESS\\
					NYU Tandon School of Engineering\\
					Brooklyn, NY 11201\\
					\{ojask, hiteshp, ychou, dshakya, shao, tsr\}@nyu.edu}}

\maketitle

\thispagestyle{firststyle}
\begin{abstract}
Future wireless cellular networks will utilize millimeter-wave and sub-THz frequencies and deploy small-cell base stations to achieve data rates on the order of hundreds of gigabits per second per user. The move to sub-THz frequencies will require attention to sustainability and reduction of power whenever possible to reduce the carbon footprint while maintaining adequate battery life for the massive number of resource-constrained devices to be deployed. This article analyzes power consumption of future wireless networks using a new metric, a figure of merit called the power waste factor ($ W $), which shows promise for the study and development of ``green G" - green technology for future wireless networks. Using $ W $, power efficiency can be considered by quantifying the power wasted by all devices on a signal path in a cascade. We then show that the consumption efficiency factor ($CEF$), defined as the ratio of the maximum data rate achieved to the total power consumed, is a novel and powerful measure of power efficiency which shows that less energy per bit is expended as the cell size shrinks and carrier frequency and channel bandwidth increase. Our findings offer a standard approach to calculating and comparing power consumption and energy efficiency.

\end{abstract}

\section{Introduction}\label{Introduction}
With the rapid expansion of 5G networks worldwide, researchers have now started focusing on attaining the promises of 6G: ultra-wide bandwidths spanning gigahertz of spectrum at carrier frequencies above 100 GHz with peak data rates up to 1 Tb/s in the 2030s, yielding an ultra-reliable network with microsecond latency and ubiquitous connectivity of conventional cellular users and a massive number of connected Internet of Things (IoT) devices \cite{akhtar2020shift, Rappaport_Access_19, Rappaport_IEEE_Spectrum_21}.

A reduction in network latency from 1 ms to microseconds may be achieved by reducing the symbol duration, increasing the computational power of processors in the core network and on edge devices, and with edge computing handling wider bandwidth channels \cite{akhtar2020shift,Rappaport_Access_19}. The connection density of devices is expected to increase from 1 million connections to 10 million connections per square kilometer in 6G \cite{shahraki2021connectivity}, resulting in greater power consumption at base stations (BSs). Improving energy efficiency for 6G networks is hence a critical problem, inherent to all aspects of design and rollout, that must be addressed \cite{NextG_Whitepaper}.

Achieving energy efficiency in future wireless devices (in particular for resource-limited IoT devices with limited battery) requires novel energy management techniques. Deep learning of network traffic patterns is one such approach being explored to optimize network resource management algorithms \cite{akhtar2020shift}. Energy harvesting, wherein the device batteries are recharged by incoming radio frequency (RF) signals, will also reduce the energy demand and extend the battery life of IoT devices \cite{rf_harvesting}. However, we are not aware of a standard figure of merit for understanding power consumption across all network components or tasks. Here, we propose a new framework to show how the cell size impacts the overall network power consumption.

Reducing the power requirements (and hence electricity consumption) for BSs will yield economic savings for network operators. More importantly, reducing the power consumed by wireless networks has significant environmental benefits - telecommunications accounts for 2 to 3 percent of total global energy demand today, but will be over 10 percent of global consumption by 2030, and a reduction in power consumed by wireless networks will lead to a reduction in greenhouse gas emissions \cite{NextG_Whitepaper}. Engineers must juggle the task of reducing the power consumed by the device with attaining gigabit-per-second data rates for future 6G applications.

Engineers may utilize new figures of merit called the \textit{power waste factor} ($ W $), the power waste figure ($W$ in dB units), and the \textit{consumption efficiency factor} ($ CEF $)  \cite{Murdock_2014} to design ``\textit{green G}" - green technology for future wireless networks\cite{NextG_Whitepaper}, by analyzing the performance trade-off (power vs. data rate) of wireless communication devices as well as network architectures in a very general way, akin to how noise is analyzed in communication systems. We define $ W $ as the ratio of the total signal path power consumed by a single device or a cascaded network divided by the power contained in the information signal at the output of the cascade (equal to the reciprocal of the power efficiency factor, $ H $, as originally defined in \cite{Murdock_2014}). The power waste factor, $W$, is always greater than or equal to unity, whereas the power efficiency factor, H, the reciprocal of W, is always less than or equal to unity \cite{Murdock_2014}. As shown here, this approach makes the analysis of wasted power on a cascade virtually identical to the analysis of additive noise along a cascade (e.g., like noise factor and noise figure). The $ CEF $ of a communication device or system, defined as the ratio of the maximum data rate achievable by the device or system to the total power consumed, can be derived for any device or linear cascaded communication system, from $ W $ \cite{Murdock_2014}. The higher the $ CEF $ of the network, the fewer Joules of energy are consumed by the network to transmit/receive one bit of information. The $ CEF $ can be derived for any general cascade communication system, making it an easy task to analyze overall power consumption and energy efficiency of any network.

The energy efficiency of individual components such as mmWave and sub-THz transmitters (TXs) and receivers (RXs) has been analyzed in prior work \cite{Murdock_2014, TX_power_subTHz,Notis_2020}; however, in this article we demonstrate how $ W $ and $ CEF $ may be used to determine and compare the energy efficiency of complete end-to-end millimeter-wave (mmWave) and sub-THz communication systems. 

In this article, we compare the $ CEF $ of a typical mmWave wireless system operating at 28 GHz to a potential 140 GHz system being developed for future 6G communication networks and show how the increased data rate at sub-THz frequencies offsets the additional power consumption of today's inefficient THz electronics.
\section{Effect of Carrier Frequency on Received Power}
When considering only free-space propagation with omnidirectional antennas, wireless channels at higher frequencies experience greater path loss due to the first meter of free space propagation loss, thus requiring greater power consumption (e.g., greater RF power at higher frequencies) at the omni TX to attain an identical signal-to-noise ratio (SNR) at the omni receiver over an identical bandwidth \cite{Rappaport_Access_19}. However, since antenna elements are typically placed a half-wavelength apart and higher frequencies result in shorter wavelengths, a greater number of antenna elements fit in an identical physical aperture area at higher frequencies. This implies that for fixed antenna area, greater gain is possible at higher frequencies, which more than overcomes the overall and first meter propagation loss at higher frequencies \cite{Rappaport_Access_19}. This fact is vital (see \cite{Rappaport_IEEE_Spectrum_21}) for destroying the myth that higher frequencies have greater path loss \cite{Rappaport_Access_19}, as small cells are vital solely because wider RF bandwidths induce more noise power, thus requiring closer BS transmitters to maintain a particular SNR for a given transmitter power, regardless of carrier frequency. As seen in \cite{Yunchou_Letter_UMi}, the omnidirectional path loss is independent of frequency in all practical indoor and outdoor channels from 28 to 140 GHz when referenced to the free space loss in the first meter of propagation. Directional antennas on each end of the link more than compensate for the frequency-dependent path loss of omnidirectional antennas in the first meter of propagation \cite{Rappaport_Access_19,  Yunchou_Letter_UMi}.

Assuming directional antennas at both ends of a radio link with constant physical antenna areas over frequency, the free space path loss \textit{decreases} quadratically as frequency increases \cite{Rappaport_Access_19,Yunchou_Letter_UMi}. Thus, for a given bandwidth, the SNR \textit{increases} with an increase in carrier frequency and fixed transmit power.

\section{Power Consumption Analysis of 28 GHz and 140 GHz Wireless Transceivers }

With the move to higher sub-THz frequencies where wider swaths of spectrum are available, the total power consumed by wireless systems typically increases due to increased system ohmic and interconnect losses and reduction in device efficiencies. However, due to the greater data rate achievable at higher frequencies, the total energy consumed by the system per bit of data transmitted \textit{decreases} \cite{Murdock_2014}. To quantify the decrease in per-bit energy consumed by the entire communication system when moving to sub-THz, $ W $ and $ CEF $ may be used. 

\subsection{Utilizing Waste Factor to Calculate Power Consumed}
The total power consumed ($P_{consumed}$) by any device or general end-to-end communication network (including computational, display, and ancillary network components) can be represented as the sum of three power components:
 
 \begin{align}
     P_{consumed} = \underbrace{P_{sig} + P_{non-sig}}_{\text{Components on signal path}}+\underbrace{P_{non-path}}_{\text{Auxiliary components}},\label{components}
 \end{align}
where $P_{sig}$ is the total output signal power delivered by all components along a cascade of components in the communication network, $P_{non-sig}$ is the total power consumed by devices on the signal path cascade that is not included in the signal output of the device (e.g., the power used or ``wasted" to support the signal transmission to the next component in the cascade), and $P_{non-path}$ is the power consumed by all ancillary components that do not carry the information signal along the cascade and which are off the signal path \cite{Murdock_2014}. Note that the total power consumed by all network components on the signal path cascade (through which the information signal flows) is equal to the sum of $P_{non-sig}$ and $P_{sig}$ \cite{Murdock_2014}.

For example, for a power amplifier that carries an information signal, $P_{sig}$ is the power at the output of the amplifier, while $P_{non-sig}$ is the power required to bias and operate the amplifier and may be considered waste since it is not directly applied to the carried signal power to the next stage. $P_{non-path}$ of the amplifier may be the power consumed by the cooling fans, and all such power values may be averaged by integrating over time or operational states.

$ W $ is an elegant and general way to quantify the total power consumed (e.g., both $ P_{sig} $ and $ P_{non-sig} $) by a communication device or a cascaded system of information-carrying components in terms of the gains and efficiencies of the devices along the signal path \cite{Murdock_2014}. $W$ is the proportion of wasted power for all components on the signal path to the useful signal power delivered at the output of the cascaded communication system, where $ W $ is the same as $ H^{-1} $ in \cite{Murdock_2014}. The total consumed power of any device or end-to-end network may be given by \eqref{H_def}, where $P_{sig}$ is the output signal power from the cascade and $ W $ denotes the added power due to waste at the signal output when referred to the input signal of the cascade. $W$ is always greater than or equal to 1, where $W=1$ denotes that all supplied power to a cascaded component or network is contributed in the signal output (optimum, no wasted power) and $W$ = $\infty$ denotes that no power is provided in the signal output and all power is wasted (e.g., an ideal dummy load or completely lossy channel). $W$ of any component or cascade is the inverse of power efficiency $H^{-1}$ for components carrying the information signal in a cascade \cite{Murdock_2014}.
\begin{align}
 P_{consumed} = (P_{sig}\times W)+ P_{non-path},\label{H_def}
\end{align}
where $ W $ for a cascade is computed by \eqref{H_formula}. 

\begin{figure}
	\centering
	\includegraphics[width=0.5\textwidth]{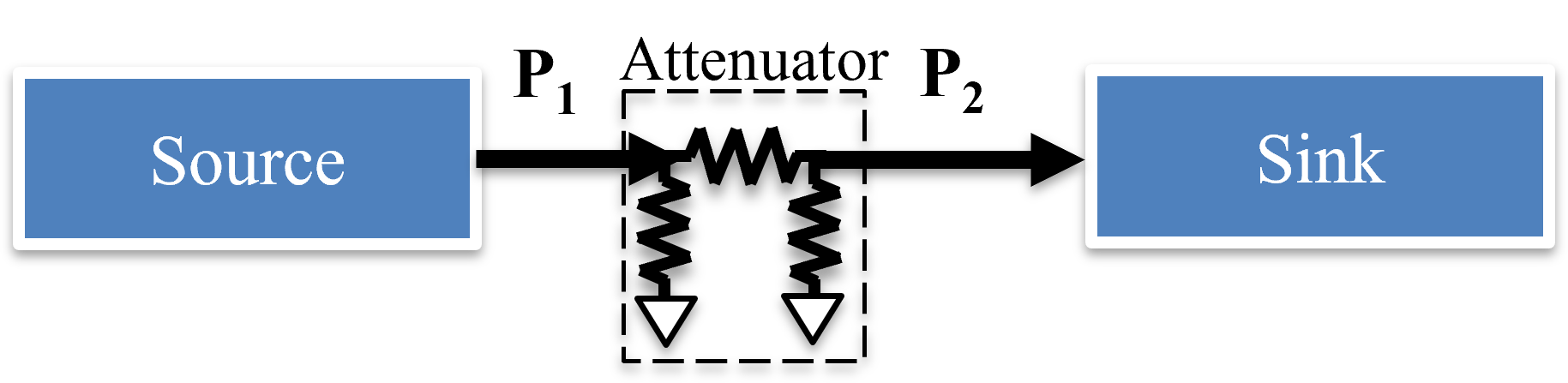}
	\caption{For a passive attenuator, $ W $ is equal to the insertion loss of the attenuator \cite{Murdock_2014}.}
	\label{fig:Passive_Load}
\end{figure}

Consider a passive attenuator connecting a source and sink, as shown in Fig. \ref{fig:Passive_Load}.
The total power consumed by the system is equal to the signal power provided by the source ($P_1$), while $P_{sig}$ is equal to the power at the output of the attenuator ($P_2$). Thus, from \eqref{H_def}, with $ P{non-path} $ = 0, $W$ of the attenuator is equal to the ratio of $P_1$ and $P_2$, that is, the insertion loss of the attenuator ($ L $), a value greater than one. For any passive device or cascade including propagation channels, $ W $ is equal to the loss ($ L $) of the device or cascade. 

As shown in \cite{Murdock_2014}, $W$,  ($ H^{-1} $ in \cite{Murdock_2014}) along a cascade is computed as:
   \begin{align}
    &W =H^{-1}= \Bigg\{W_N+\dfrac{\left(W_{N-1}-1\right)}{G_N}\nonumber\\
    &+\dfrac{\left(W_{N-2}-1\right)}{G_NG_{N-1}}+\dotsc+\dfrac{\left(W_1-1\right)}{\prod_{i = 2}^NG_i}\Bigg\}\label{H_formula},
\end{align}
 where the first component in the cascade is closest to the message source and the Nth component in the cascade is closest to the sink, $W_i$ is the power waste factor, and $G_i$ is the gain of each component along the cascade. 

To reduce the power consumed in a wireless communication system, engineers must minimize $W$ by optimizing the gains and efficiencies of components on the signal path, with clear advantages gained by making some components, such as the final amplifier stages, less wasteful than others, as seen from \eqref{H_formula}. Ancillary components not on the signal path are not considered in calculating $ W $ but are added at the end to determine total power consumption, as shown in \eqref{H_def} \cite{Murdock_2014}.

Eq. \eqref{H_def} and \eqref{H_formula} show that the total power consumed by a wireless communication device, system, or cascade may be computed by separately evaluating $ W $ and $ G $ for individual components and then using \eqref{H_formula}

It becomes clear from inspection that \eqref{H_formula} is analogous to the computation of noise figure (NF), where $W$ (wasted power) is somewhat analogous to the noise factor, which indicates the amount of noise power related to the input of a device or cascade. Here, $ W $ is referenced to the output, not the input as in noise figure, and 10 log $ W $ is defined as the power waste figure ($ W $(dB)).  That is, as shown in \cite{Murdock_2014}, NF = 0 dB indicates there is no additive noise in a cascade and there is no degradation in SNR, and quite analogously, $W$ = 0 dB indicates all power consumed in a cascade is contained in the signal output and no power is wasted. Splitting the total power consumed by any communication system into three components as in \eqref{components} facilitates the derivation of general expressions to calculate the Joules consumed by a communication system to transmit a single bit of data, using $H^{-1}= W $ \cite{Murdock_2014}.

\subsection{$ CEF $ Theory for Power Consumption Analysis}
\begin{figure}
	\centering
	\includegraphics[width=0.5\textwidth]{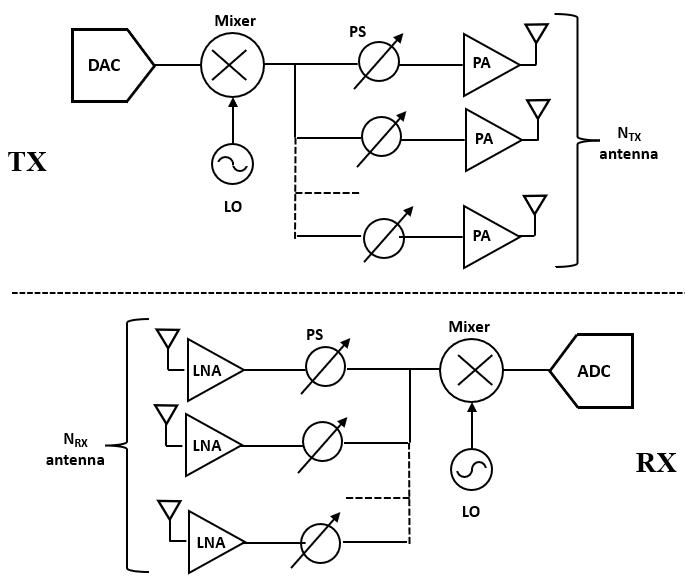}
	\caption{Phased array antenna architecture at the TX and RX with direct-conversion considered for power analysis.}
	\label{fig:TRX_arch}
\end{figure}

The consumption efficiency factor ($CEF$) is defined as the ratio of the maximum data rate supported by a communications system to the total power consumed by the system \cite{Murdock_2014}. $ CEF $ was first introduced as \textit{consumption factor} ($ CF $) in \cite{Murdock_2014} and is renamed here to stress the fact that $CEF$ is the energy \textit{efficiency} of the communication system in terms of energy per bit. The $CEF$ quantifies the number of bits a communication system may transmit per Joule of energy consumed. Systems with lower per-bit energy consumption (i.e., higher $CEF$) are preferred from the perspective of overall green G energy efficiency, with the upper limit of $ CEF $ (as the bandwidth approaches infinity) derived using Shannon's formula (see (72) in \cite{Murdock_2014}).

The effect of varying communication system parameters, such as system bandwidth, device efficiencies, carrier frequency, and cell radius, can be quickly evaluated by evaluating $ W $ and $ CEF $, allowing engineers to compare competing deployments or products or simple circuit architectures with standard metrics steeped in fundamentals.

\subsection{System Parameters $ W $ and $ CEF $ Analysis}\label{sec:system_parameters}
We now compare $ W $ and $ CEF $ of two communication systems - an mmWave system operating at 28 GHz with an RF bandwidth of 400 MHz (equal to the maximum bandwidth for FR2 in the 3rd Generation Partnership Project, 3GPP, standard) and a sub-THz system with an RF bandwidth of 4 GHz, operating at 140 GHz.  To analyze transceiver power consumption, we assume analog beamforming (at the BS and user equipment, UE), as used by systems today, as illustrated in Fig. \ref{fig:TRX_arch}.

For both the mmWave and sub-THz systems, the BS-UE distance is 100 m, a path loss exponent (PLE) of 2.0 was assumed for line of sight (LoS) propagation and a PLE of 3.2 was assumed in non- LoS (NLoS) environments, as obtained from the multi-band path loss close-in (CI) reference distance path loss model with $ d_0 $ = 1 m as in \cite{Yunchou_Letter_UMi} for Urban Microcell (UMi) propagation. The effective antenna aperture area at the BS and the UE was kept constant at 0.5 m$^2$ and 5 cm$^2$, respectively, resulting in antenna gains of 45.2 and 15.2 dBi at 28 GHz and antenna gains of 59.1 and 29.1 dBi at 140 GHz, respectively (assuming antenna efficiencies of 0.6). These gains correspond to number of BS antenna elements of 1024 and 4096 at 28 and 140 GHz, respectively, while fewer antenna elements were assumed to be present at the UE due to less aperture area - eight elements at 28 GHz and 64 elements at 140 GHz. The power efficiency of the TX power amplifier was assumed to be 28 percent at 28 GHz as per specifications of the CMD262 power amplifier \cite{CMD262}; however, a lower efficiency of 20.8 percent was assumed for the power amplifier at 140 GHz \cite{subTHz_PA_efficiency}. To calculate the power consumed by the many low-noise amplifiers (LNAs), we used the Figure of Merit (FoM) of an LNA that quantifies the DC power drawn by the LNA \cite{Notis_2020}. Energy efficient LNAs have high FoM. The FoM of the LNAs were found to be 24.83 mW$^{-1}$ and 8.33 mW$^{-1}$ at 28 and 140 GHz, respectively \cite{LNA_28_FoM, LNA_140_FoM}. Since the DC power drawn by the LNA at either BS or UE is independent of the signal power at the output of the LNA, the power drawn by the LNA is added as non-path power to the overall power consumed by the system in \mbox{\eqref{H_def}}, while $ W $ for the LNA is set to be equal to 1, that is, the LNA is modeled to consume no signal-path power but is treated with known gain and fixed auxiliary power drain. The gain of the LNA was assumed to be 20 dB at both frequencies. In general, device components at sub-THz frequencies are less efficient since the technology is not as mature as mmWave devices. The mixers and phase shifters were assumed to be passive devices, with an insertion loss of 6 dB and 10 dB, respectively, in both bands \cite{Notis_2020}. The power consumed by the non-path local oscillator (LO) was assumed to be 10 dBm and 19.9 dBm at 28 GHz and 140 GHz, respectively\cite{Notis_2020}.

In addition to the power consumed by the RF components of the BS and UE, there is a power overhead required to keep the circuitry cool, and this is considered to be non-path power in \eqref{H_def}. As per \cite{Arnold_2010}, a 20 percent cooling overhead is assumed at the BS. Since UEs must be portable, passive cooling is utilized (which consumes no energy), wherein heat is dissipated by thermally conductive material. Additionally, we assume that UE screens consume 500 mW of power.

\section{Numerical Results}
Using the system parameters listed earlier, assuming a transmit power of 1 mW, $ W $ of the end-to-end system is calculated via \eqref{H_formula}. The total power consumed by the end-to-end system ($P_c$) is then calculated via \eqref{H_def}. The data rate of the system ($R$) is approximated via Shannon's formula (using (45) in \cite{Murdock_2014}), while the $CEF$ is equal to the ratio of $R$ and $P_c$ (see (44) in \cite{Murdock_2014}).
\subsection{Comparison of the energy efficiency of the mmWave and sub-THz Systems}
Using \eqref{H_formula}, for uplink transmissions, the 28 GHz mmWave end-to-end system (composed of the UE TX, the mmWave channel, and the BS RX) has $ W $ of 52.2 dB for LoS environments and 76.2 dB for NLoS environments, while the 140 GHz sub-THz system (composed of the UE as the TX, the sub-THz channel and the BS as the RX) achieves a more efficient $W$ of 48.0 dB and 72.0 dB for LoS and NLoS environments, respectively. The large values of $W$ are due to the large channel path loss and high power consumption of the mmWave and sub-THz systems, with the power amplifiers consuming the most power along the signal path (note: channels are also considered as being part of the cascade \cite{Murdock_2014}, and are included in the computation of $W$). The wireless channel itself acts as a passive attenuator, attenuating the RF signal propagating through the channel, but consumes no power. The wireless channel can thus be incorporated into \eqref{H_def} and \eqref{H_formula} as a simple attenuator, with $ W $ equal to the path loss of the channel. For downlink data transmission, the $W$ of the end-to-end mmWave system is 72.7 dB and 96.7 dB for LoS and NLoS environments, respectively, while the $W$ of the end-to-end sub-THz system is 66.0 dB and 90.0 dB for LoS and NLoS environments, respectively. Note that $W$ of the communication systems (e.g., 72.7 dB for downlink transmission for LoS environments at 28 GHz) is less than the total link path loss (e.g., 101.4 dB for LoS environments at 28 GHz) as the link path loss is compensated for by the gain of the TX and RX antennas and power amplifiers (see (75) in \cite{Murdock_2014}). At a fixed BS-UE separation distance of 100 m, NLoS environments have smaller $ CEF $ than LoS channels since lower data rate is achieved due to lower SNR (caused by greater path loss \cite{Yunchou_Letter_UMi}) at the RX. The comparison of the 28 and 140 GHz systems due to received power ($P_r$), $R$, $W$, $P_c$, and $CEF$ are summarized in Table \ref{tbl:CF}.

At higher frequencies, due to a greater number of power amplifiers and LNAs required for the additional phased array antenna elements, the total power consumed by the end-to-end system (composed of the UE, the wireless channel, and the BS) by the BS and UE increases. Additionally, the power consumed by the analog-to-digital converter (ADC) and digital-to-analog converter (DAC) increases in proportion to the bandwidth of operation \cite{Notis_2020}. Using \eqref{H_def}, the net power consumed by the end-to-end system (composed of the UE, the wireless channel, and the BS) increased from 4.95 W to 24.88 W for downlink LoS transmissions when moving from 28 to 140 GHz, and from 6.59 to 40.07 W for uplink transmissions when moving from 28 to 140 GHz. \textit{However, the increase in raw circuit power consumption is compensated for by a greater increase in data rate at 140 GHz provided by the wider RF bandwidth channel, leading to a greater $ CEF $ and improved energy efficiency per bit.}

\begin{table*}
\centering
\caption{Variation in the free space path loss, received power, data rate, $W$, and $ CEF $ for the 28 and 140 GHz systems.}
\label{tbl:CF}
\resizebox{\textwidth}{!}{
\begin{tabular}{|c|c|cc|clll|clll|clll|clll|c|}
\hline
\multirow{3}{*}{\textbf{\begin{tabular}[c]{@{}c@{}}fc\\ (GHz)\end{tabular}}} & \multirow{3}{*}{\textbf{\begin{tabular}[c]{@{}c@{}}FSPL(fc, 1 m)\\ (dB)\end{tabular}}} & \multicolumn{2}{c|}{\textbf{\begin{tabular}[c]{@{}c@{}}$P_r$\\(dBW)\end{tabular}}}&\multicolumn{4}{c|}{\textbf{\begin{tabular}[c]{@{}c@{}}$W$\\(dB)\end{tabular}}}&\multicolumn{4}{c|}{\textbf{\begin{tabular}[c]{@{}c@{}}$P_c$\\(Watts)\end{tabular}}}&\multicolumn{4}{c|}{\textbf{\begin{tabular}[c]{@{}c@{}}R\\(Gbps)\end{tabular}}}&\multicolumn{4}{c|}{\textbf{\begin{tabular}[c]{@{}c@{}}$ CEF $\\(Gb/J)\end{tabular}}}&\multirow{3}{*}{\textbf{Comments}}\\\cline{3-20}
&&\multicolumn{1}{c|}{\multirow{2}{*}{LoS}}&\multirow{2}{*}{NLoS}&\multicolumn{2}{c|}{LoS}&\multicolumn{2}{c|}{NLoS}&\multicolumn{2}{c|}{LoS}&\multicolumn{2}{c|}{NLoS}&\multicolumn{2}{c|}{\multirow{2}{*}{LoS}}&\multicolumn{2}{c|}{\multirow{2}{*}{NLoS}} & \multicolumn{2}{c|}{LoS}& \multicolumn{2}{c|}{NLoS}&\\ \cline{5-12} \cline{17-20}
&&\multicolumn{1}{c|}{}&&\multicolumn{1}{c|}{UL}&\multicolumn{1}{c|}{DL}&\multicolumn{1}{c|}{UL}&\multicolumn{1}{c|}{DL}&\multicolumn{1}{c|}{UL}&\multicolumn{1}{c|}{DL}&\multicolumn{1}{c|}{UL}&\multicolumn{1}{c|}{DL}&\multicolumn{2}{c|}{}&\multicolumn{2}{c|}{}&\multicolumn{1}{c|}{UL}&\multicolumn{1}{c|}{DL}&\multicolumn{1}{c|}{UL}&\multicolumn{1}{c|}{DL}& \\ \hline
\textbf{28}&\textbf{61.4}&\multicolumn{1}{c|}{\textbf{-71.1}}&\textbf{-95.1}&\multicolumn{1}{c|}{\textbf{52.2}}&\multicolumn{1}{l|}{\textbf{72.7}}&\multicolumn{1}{l|}{\textbf{76.2}}& \textbf{96.7}&\multicolumn{1}{c|}{\textbf{6.59}}&\multicolumn{1}{l|}{\textbf{4.95}}&\multicolumn{1}{l|}{\textbf{6.59}}& \textbf{4.95}&\multicolumn{2}{l|}{\textbf{4.89}}& \multicolumn{2}{l|}{\textbf{1.73}}& \multicolumn{1}{l|}{\textbf{0.74}}  & \multicolumn{1}{l|}{\textbf{0.99}} &\multicolumn{1}{l|}{\textbf{0.26}} & \textbf{0.35}& \textbf{\begin{tabular}[l]{@{}l@{}} \tabitem Lower power consumption\\ \tabitem Lower $ CEF $.\\\tabitem Less energy efficient per bit\end{tabular}}\\ \hline
\textbf{140}&\textbf{75.4}&\multicolumn{1}{c|}{\textbf{-57.1}}&\textbf{-81.1}&\multicolumn{1}{c|}{\textbf{48.0}}&\multicolumn{1}{l|}{\textbf{66.0}}&\multicolumn{1}{l|}{\textbf{72.0}}&\textbf{90.0}&\multicolumn{1}{c|}{\textbf{40.07}}&\multicolumn{1}{l|}{\textbf{24.88}}&\multicolumn{1}{l|}{\textbf{40.07}}&\textbf{24.88}&\multicolumn{2}{l|}{\textbf{54.16}}&\multicolumn{2}{l|}{\textbf{22.39}}&\multicolumn{1}{l|}{\textbf{1.35}}&\multicolumn{1}{l|}{\textbf{2.18}}&\multicolumn{1}{l|}{\textbf{0.56}}&\textbf{0.90}& \textbf{\begin{tabular}[l]{@{}l@{}} \tabitem Higher data rate \\ \tabitem Higher power consumption\\ \tabitem Higher $ CEF $. \\ \tabitem More energy efficient per bit\end{tabular}}\\\hline
\end{tabular}
}
\end{table*}
\subsection{Effect of System Bandwidth on $ CEF $}
The vast spectrum available at sub-THz frequencies allows wireless systems to operate with wide bandwidths and achieve data rates on the order of hundreds of gigabits per second per user. Thermal noise increases in proportion to the system bandwidth, requiring greater effective isotropic radiated power (EIRP) for an RX to attain the same SNR as smaller bandwidth channels. We now investigate the system bandwidth at which it may be advantageous to switch to higher-frequency systems from an optimal ``bits per Joule" energy consumption perspective. 

To observe the effect of bandwidth on $ CEF $, the bandwidth of the mmWave system is kept constant at 400 MHz, while the bandwidth of the sub-THz system is varied from 100 MHz to 8 GHz. The total power consumed by the wireless system is calculated using \eqref{H_def} and $ W $ of the system, and the data rate is calculated via (45) in \cite{Murdock_2014}. A BS-UE link distance of 100 m is assumed.

\begin{figure}
	\centering
\subfigure[The variation in the $ CEF $ of the sub-THz system with bandwidth for uplink transmissions.]{
		\includegraphics[width=0.4\textwidth]{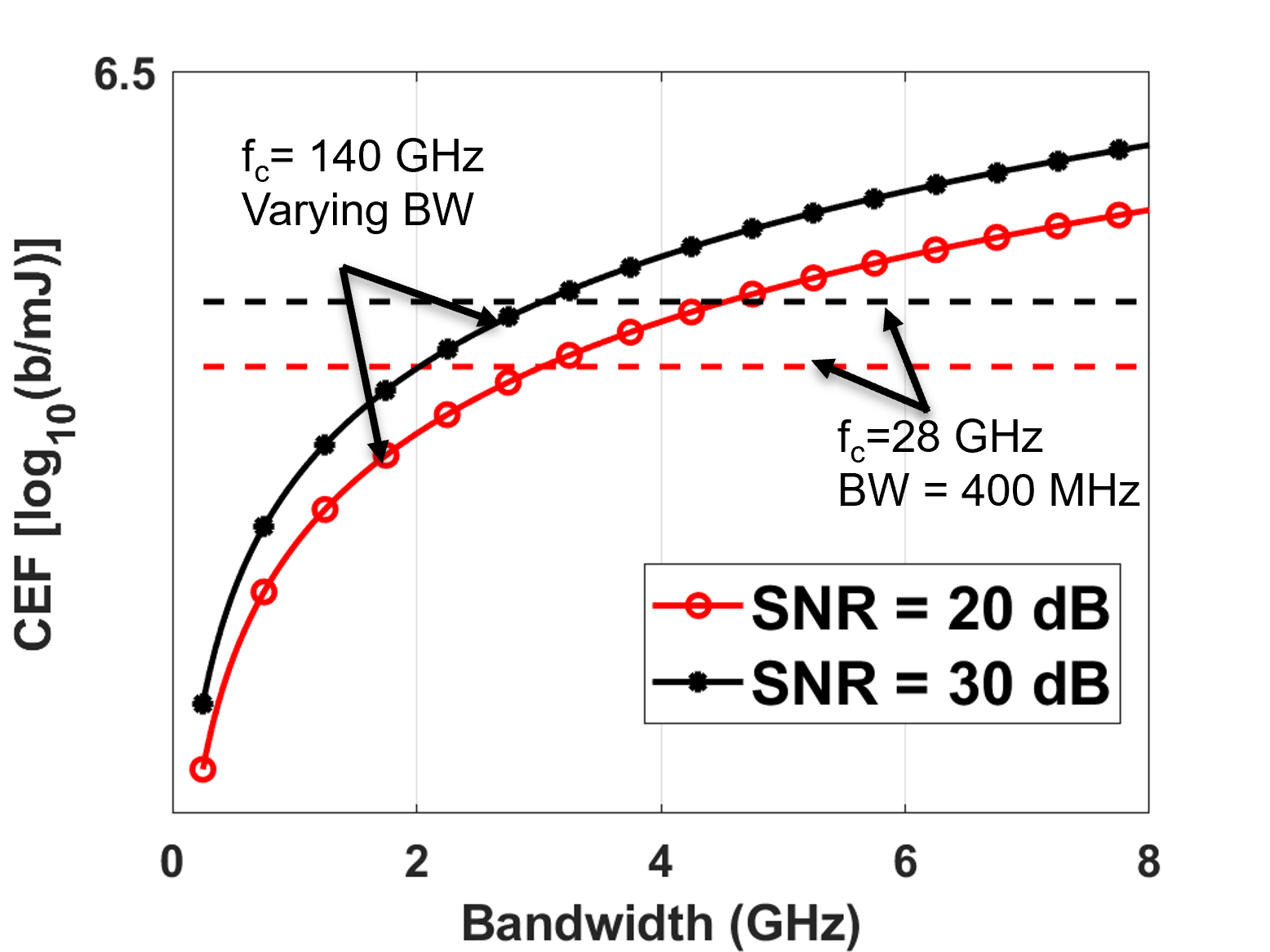}
		\label{fig:CF_BW_SNR_UL}}
	\subfigure[The variation in the $ CEF $ of the sub-THz system with bandwidth for downlink transmissions.]{
		\includegraphics[width=0.4\textwidth]{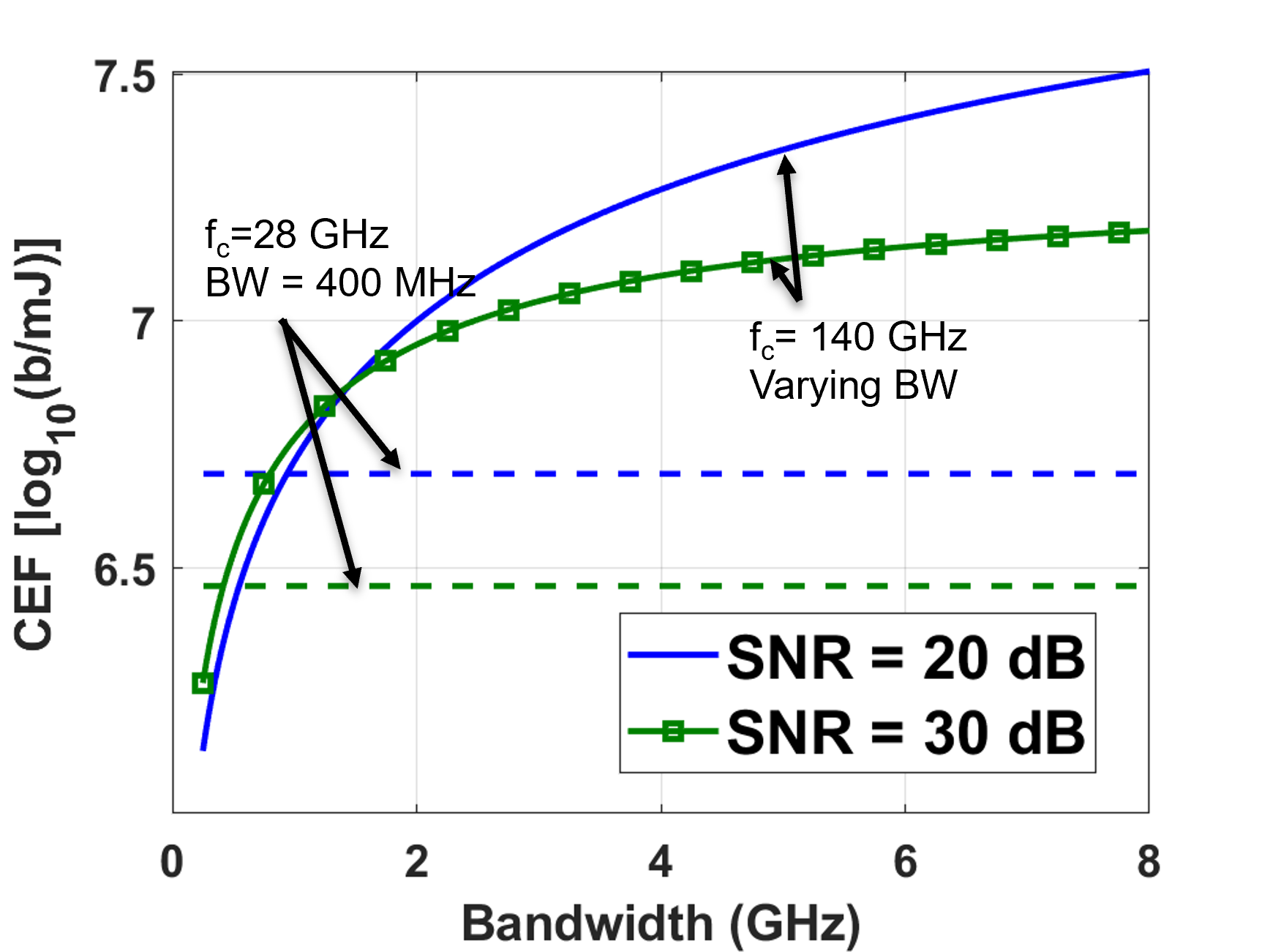}
		\label{fig:CF_BW_SNR_DL}}
	\caption{The $ CEF $ of the sub-THz system is greater than the mmWave system even at bandwidths as low as 1 GHz.}
		\label{fig:CF_BW_SNR}
\end{figure}

We observe a threshold bandwidth for uplink and downlink transmissions, above which the $ CEF $ for the sub-THz system is greater than the $ CEF $ of the mmWave system. As is seen in Fig. \ref{fig:CF_BW_SNR_DL}, for downlink transmissions at an SNR of 20 dB, the $ CEF $ of the sub-THz system with a bandwidth greater than 950 MHz is greater than the $ CEF $ of the mmWave system operating at 400 MHz. Furthermore, a crossover point was observed between the $ CEF $ curves with SNR of 20 dB and 30 dB at a bandwidth of 1 GHz. Maintaining a lower SNR at wider bandwidths may be more energy-efficient since the increase in device power consumption with bandwidth and the increase in required EIRP to maintain a constant SNR dominates the increased data rate due to higher SNR. 

For uplink transmissions at an SNR of 20 dB, while keeping the bandwidth of the 28 GHz mmWave system fixed at 400 MHz, as seen in Fig. \ref{fig:CF_BW_SNR_UL}, if the bandwidth of the sub-THz 140 GHz system is greater than 3.25 GHz (roughly eight times the bandwidth of the mmWave system), the $ CEF $ of the sub-THz system is greater than the $ CEF $ of the mmWave system. Thus, appropriate spectral allocation, guided by $ CEF $ analysis, is vital for energy-efficient future sub-THz wireless systems.

The thresholds on the uplink and downlink bandwidth mentioned above depend on the power efficiencies, gains, and power consumption models of the signal path devices and ancillary components not on the signal path of the mmWave and sub-THz systems, and may be compared using (1) - (3) here, based on details of the wireless systems.
\subsection{Effect of Component Efficiency on $ CEF $}
\begin{figure}
	\centering
	\includegraphics[width=0.5\textwidth]{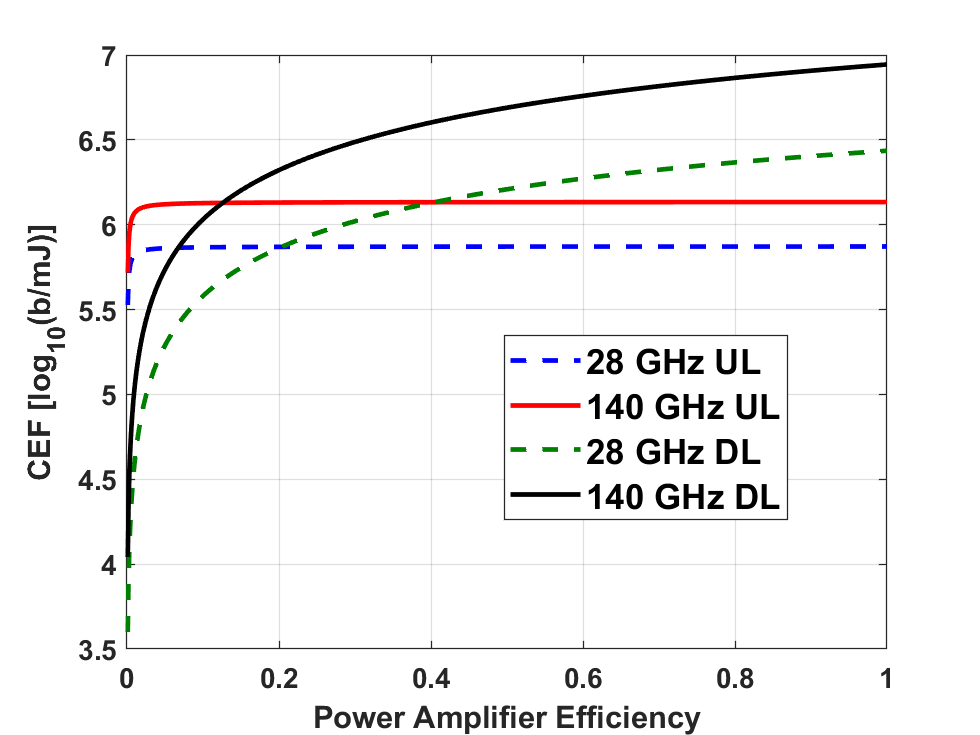}
	\caption{The effect of the power amplifier efficiency on the $ CEF $. When using much wider RF channel bandwidths at sub-THz, to attain the same $ CEF $ as the mmWave system, the required efficiency of power amplifiers of the sub-THz system is lower.}
	\label{fig:PA_eff}
\end{figure}

The power waste factor $W$ of signal-path components such as the power amplifiers provides a measure of the power ``wasted," that is, the power used by the component which is not directly converted to RF power \cite{Murdock_2014}. Since sub-THz circuits are currently in early stages of development, the power efficiency of sub-THz power amplifiers is typically lower than the efficiency of mmWave power amplifiers. However, since sub-THz systems require less TX EIRP to attain identical SNR as mmWave systems (due to greater antenna gains for identical antenna aperture areas \cite{Rappaport_Access_19}), less RF power will be needed in power amplifiers at sub-THz frequencies. For example, the effect of varying the power amplifier efficiency on the $ CEF $ is depicted in Fig. \ref{fig:PA_eff}, for uplink and downlink transmissions of the mmWave and sub-THz systems with bandwidths of 400 MHz and 4 GHz, respectively, with a BS-UE link distance of 100 m. To attain the same downlink $ CEF $ of an mmWave system with a power amplifier efficiency of 0.2 (e.g. a $ CEF $ of 0.73 Gb/J as found by using (1)-(3)), the minimum required efficiency of the sub-THz power amplifier is only 0.07.

For downlink transmissions, since the BS possesses a greater number of amplifiers, the effect of varying the amplifier efficiency is more pronounced, whereas in uplink transmissions, the $ CEF $ is insensitive to power amplifier efficiency, since the LNA at the BS consumes a majority of the power consumed by the wireless system. Insights vital for green-G were provided in \cite{Murdock_2014}, such as the need for the most efficient devices to be closest to the sink of a network, the need for directional antennas, and appropriate use of relays, repeaters, or intelligent reflecting surfaces (IRS). In both uplink and downlink transmissions, the $ CEF $ at 140 GHz is greater than the $ CEF $ at 28 GHz for identical power amplifier efficiency, \textit{which supports this analysis that wireless communication systems at higher frequencies will be more energy-efficient on a per-bit basis.}

\subsection{Effect of Cell Radius on $ CEF $}
Small cell technology (i.e., network densification) is one of the three key pillars of 5G technology, in addition to the vast mmWave spectrum and massive multiple-input multiple-output (MIMO) antenna technology \cite{Rappaport_IEEE_Spectrum_21}. Since the distance between neighboring BSs decreases with the deployment of small cells, appropriate BS power control is required to ensure that inter-cell interference does not increase. For example, since the EIRP required to maintain a target SNR at the cell edge reduces by 6 dB if the cell radius is halved (assuming a PLE of 2.0), reducing the transmit power by 6 dB will ensure no increase in interference.

To analyze the effect of varying the cell radius on the $ CEF $, the total throughput and total power consumed by a sub-THz communication network spanning a 1 km$^2$ area was calculated with varying cell radius. The sub-THz network is comprises hexagonal cells with one BS at the center of each cell. The BS was assumed to have six phased antenna arrays, and 15 active UEs were distributed uniformly throughout the cell. A system bandwidth of 4 GHz was divided between the active users served by each phased array. The device parameters of the BS and UE are as listed above. The cell radius of each BS was varied from 20 m to 500 m, while a constant target SNR of 20 dB was maintained at UE in an LoS environment near the cell edge. The NYU (squared) LoS probability model was used to predict the outage/blockage environment of each UE based on the distance of the UE from the BS, with UE close to a BS more likely to be in LoS  \cite{Rappaport_Access_19}. 

\begin{figure}
	\centering
	\includegraphics[width=0.5\textwidth]{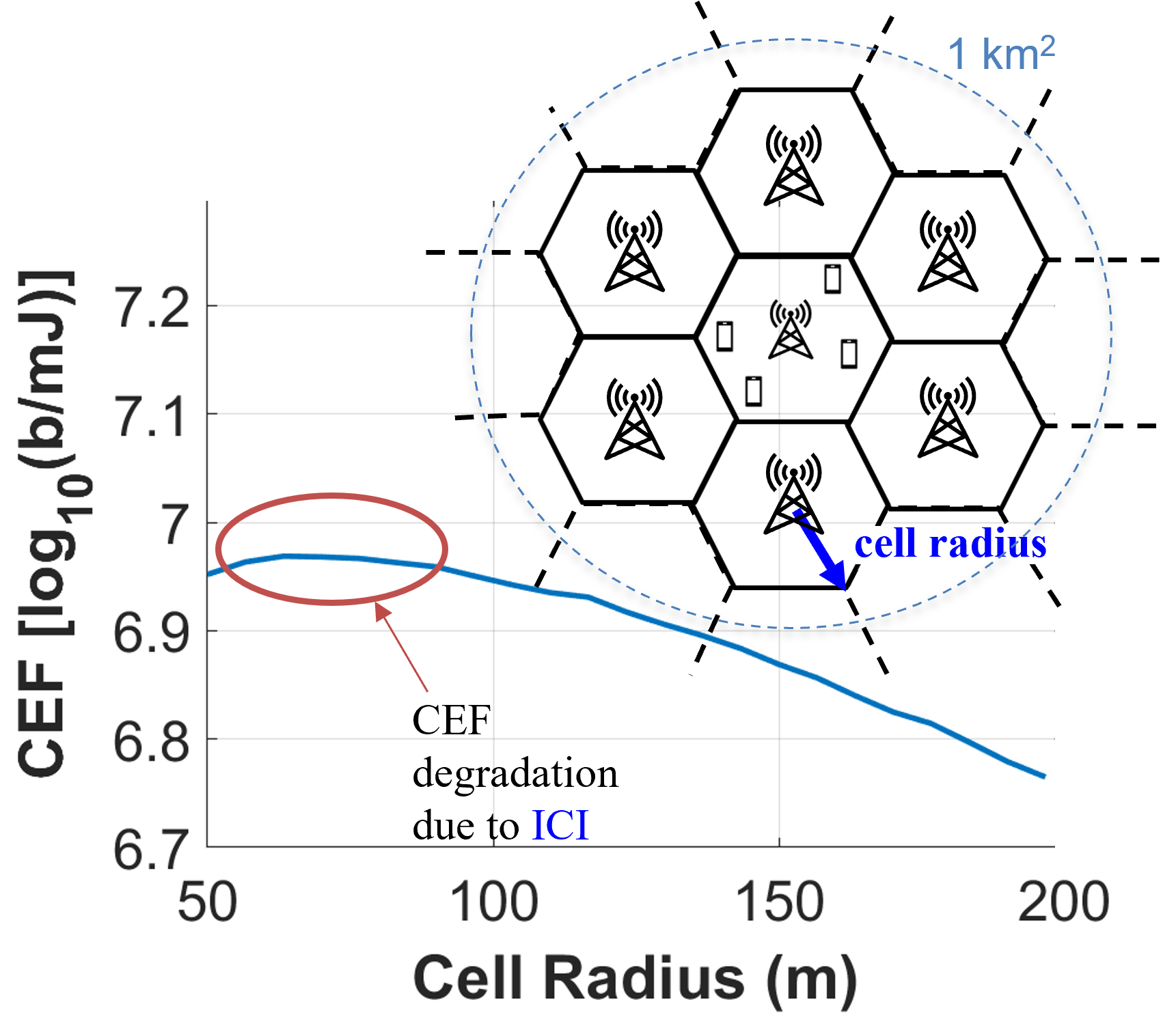}
	\caption{By using small cells in future 6G communication networks, a greater consumption efficiency factor ($ CEF $) may be attained. }
	\label{fig:small_cell}
\end{figure}

As is seen in Fig. \ref{fig:small_cell}, reducing the cell radius dramatically improves the $ CEF $ of the system, with an optimal cell radius of 65 m observed. When the cell radius was reduced below 65 m, inter-cell interference (ICI) reduces the UE SINR, reducing the maximum achievable data rate, which in turn decreases $ CEF $.

Although increasing the number of small cell BSs deployed decreases the per-bit energy cost, network operators must also take into account the economic cost of deploying additional BSs. Initial small cell deployments will target areas with greater population density for greater economic returns on investment.
\section{Future Research Directions}
 More research is warranted to see if a similar standardized mathematical framework of energy efficiency, using power superposition (\eqref{H_def}) and figures of merit such as $ W $ and $ CEF $, can be used to quantify and analyze the caching, algorithmic design, and computational energy burden of non-path devices, which dominate the power consumption of most wireless networks. The build-out of small cells will tax the global power grid; thus, designing optimal energy-efficient network architectures is vital to keep 6G green. IoT will be a big part of the future wireless ecosystems, and energy efficiency improvements will be required for those devices with limited batteries. Work on RF energy harvesting techniques at mmWave and sub-THz frequencies could close the gap by recovering a portion of energy consumed by converting incident RF power into DC energy via rectifier circuits  \cite{rf_harvesting}.

The narrower beamwidths at sub-THz frequencies result in a larger search space for optimal beam pointing directions, requiring a greater number of reference signals to be exchanged between the BS and UE, causing potential increase in power consumption. Novel beam management techniques must be explored to reduce initial access and reconnection times and power consumed \cite{akhtar2020shift}. 

The energy overhead for beam management could be reduced via the deployment of IRSs. An IRS consists of a large array of scattering elements that help direct propagating signals in the desired direction, improving the signal-to-interference-plus noise (SINR) at the RX, which results in fewer antenna elements required at the RX, reducing device complexity and power consumption \cite{akhtar2020shift}. The design of chips, transceivers, sensor architectures, as well as the performance of beam management techniques and the optimal placement of IRSs could be analyzed via $ CEF $ theory, $ W $, for engineers to attain maximal energy-efficient network design.

\section{Conclusion}\label{sec:conclusion}
This article has introduced a framework to analyze the power consumption of wireless devices and networks via two new figures of merit that denote the waste in a signal path of a device or cascade, W, and the ratio of the maximum data rate achieved by a communications system to the overall power consumption, $ CEF $. $ W $ may be used by engineers to study the impact of varying the gains and efficiencies of cascaded signal path devices on the power consumed by a device, system, or cascade. The $ CEF $ provides a quantitative metric for the trade-off between the data rate and the power consumed by a communication system using $W$. The $ CEF $ of mmWave and sub-THz wireless devices operating at 28 GHz and 140 GHz, respectively, has been compared to quantify the effect of moving to higher carrier frequencies on power consumption. For the example systems given here, if the bandwidth allocated to sub-THz systems is above a threshold bandwidth of 950 MHz for downlink transmissions and 3.25 GHz for uplink transmissions, a greater number of bits per Joule may be transmitted compared to the mmWave system operating with an RF bandwidth of 400 MHz (assuming equal antenna array aperture areas). The increase in circuit power consumption is compensated for by a greater increase in data rate. By shrinking the cell size, further improvement in energy efficiency is attainable; however, network operators must also consider the monetary expenditure required to deploy additional BS.

\section{Acknowledgments}
This material is based upon work supported by the NYU WIRELESS Industrial Affiliates Program and National Science Foundation (NSF) Grants: 1909206 and 2037845.

\section*{Biographies}
\textsc{Ojas Kanhere} received the Ph.D. degree in electrical engineering from NYU WIRELESS Research Center, New York University (NYU) Tandon School of Engineering, Brooklyn, NY, USA in 2022, under the supervision of Prof. Rappaport. He received the B.Tech. and M.Tech. degrees in Electrical Engineering from IIT Bombay, Mumbai, India, in 2017.  His research interests include mmWave and sub-THz localization, and wireless channel measurements and modeling.

\bigskip
\noindent\textsc{Hitesh Poddar} is currently pursuing a Ph.D. degree in electrical engineering with the NYU WIRELESS Research Center, New York University Tandon School of Engineering, Brooklyn, NY, USA, under the supervision of Prof. Theodore S. Rappaport. Prior to starting his Ph.D. in 2021 he worked at Qualcomm India for four years in 2G and 5G technologies. His work has been primarily focused on algorithm design, and implementation for the physical layer. He received his B.TECH from VIT, Vellore in 2017.  His research interests include millimeter-wave and Terahertz channel measurement, channel modeling, and ray tracing.

\bigskip
\noindent\textsc{Yunchou Xing} received the B.S. degree in electronic science and technology from Tianjin University, Tianjin, China, in 2014, and the M.S. degree in electrical engineering from the Tandon School of Engineering, New York University (NYU), Brooklyn, NY, USA, in 2016. He received his Ph.D. in Dec 2021, majoring in millimeter-wave (mmWave) and THz wireless communications under the supervision of Prof. Theodore. S. Rappaport. He joined NOKIA in Feb 2022. He has authored or co-authored over 20 technical papers in the field of mmWave and THz wireless communications. His research interests include machine learning, radio propagation, channel sounding, and channel modeling for ultra-wideband communications systems with a focus on frequencies above 100 GHz.

\bigskip
\noindent\textsc{Dipankar Shakya} received the M.S. degree in Electrical Engineering from New York University in 2021 and the B.E degree in Electronics and Communications from Tribhuwan University, Nepal, in 2016. He is currently pursuing a Ph.D. degree in electrical engineering with the NYU WIRELESS Research Center, New York University Tandon School of Engineering, Brooklyn, NY, USA, under the supervision of Prof. Theodore S. Rappaport. He joined the NYU WIRELESS Research Center in 2019 following three years of service as an engineer for flood early warning systems in South Asia. His research interests include millimeter-wave and Terahertz channel measurement systems and RF circuit design.

\bigskip
\noindent\textsc{Shihao Ju} received the B.S. degree in communications engineering from the Harbin Institute of Technology, Harbin, China, in 2017, and the M.S. and Ph.D. degree in electrical engineering at NYU WIRELESS, New York University (NYU), Brooklyn, NY, USA, in 2019, and 2022, respectively, under the supervision of Prof. Theodore S. Rappaport. He has authored or co-authored 18 technical papers in mmWave and THz wireless  communications and signal processing. His research interests include millimeter-wave and Terahertz propagation measurements, statistical channel modeling and simulation, reconfigurable intelligent surface, and machine learning applications on physical-layer technologies.

\bigskip
\noindent\textsc{Theodore S. Rappaport} is the David Lee/Ernst Weber Professor at New York University (NYU). He founded NYU WIRELESS and the wireless research centers at the University of Texas Austin (WNCG) and Virginia Tech (MPRG). His research has provided fundamental knowledge of wireless channels used to create the IEEE 802.11 standard, the first U.S. digital TDMA and CDMA standards, the first public Wi-Fi hotspots, and recently proved the viability of mm-wave and sub-THz frequencies for 5G, 6G, and beyond. He founded two companies that were sold to publicly traded companies – TSR Technologies, Inc. and Wireless Valley Communications, Inc., and was an advisor to Straight Path Communications which sold 5G mm-wave spectrum to Verizon.


\end{document}